\documentstyle[multicol,aps,psfig]{revtex}
\renewcommand{\narrowtext}{\begin{multicols}{2}
\global\columnwidth20.5pc\noindent}
\renewcommand{\widetext}{\end{multicols}
\global\columnwidth42.5pc}
\begin{document}
\draft
\preprint{30 August 1999}
\title{Characterization of ferrimagnetic Heisenberg chains
       according to the constituent spins}
\author{Shoji Yamamoto}
\address{Department of Physics, Okayama University,
         Tsushima, Okayama 700-8530, Japan}
\author{Takahiro Fukui}
\address{Institut f\"ur Theoretische Physik, Universit\"at zu K\"oln,
         Z\"ulpicher Strasse 77, 50937 K\"oln, Germany}
\author{T$\hat{\mbox o}$ru Sakai}
\address{Department of Physics, Himeji Institute of Technology,
         Ako, Hyogo 678-1297, Japan}
\date{Received 30 August 1999}
\maketitle
\begin{abstract}
   The low-energy structure and the thermodynamic properties of
ferrimagnetic Heisenberg chains of alternating spins $S$ and $s$ are
investigated by the use of numerical tools as well as the spin-wave
theory.
The elementary excitations are calculated through an efficient
quantum Monte Carlo technique featuring imaginary-time correlation
functions and are characterized in terms of interacting spin waves.
The thermal behavior is analyzed with particular emphasis on its
ferromagnetic and antiferromagnetic dual aspect.
The extensive numerical and analytic calculations lead to the
classification of the one-dimensional ferrimagnetic behavior
according to the constituent spins: the ferromagnetic ($S>2s$),
antiferromagnetic ($S<2s$), and balanced ($S=2s$) ferrimagnetism.
\end{abstract}

\narrowtext
\section{Introduction}\label{S:I}

   One-dimensional quantum ferrimagnets are one of the hot topics and
recent progress \cite
{Alca67,Tian53,Pati94,Breh21,Ono76,Nigg31,Ivan24,Yama10,Yama11,Yama08,Mais08,Kura62,Yama24,Wu57,Saka53}
in the theoretical understanding of them deserves special mention.
Such a vigorous argument a great deal originates in the pioneering
efforts to synthesize binuclear magnetic materials including
one-dimensional systems.
The first ferrimagnetic chain compound \cite{Glei27},
MnCu(dto)$_2$(H$_2$O)$_3$$\cdot$$4.5$H$_2$O
($\mbox{dto}=\mbox{dithiooxalato}=\mbox{S}_2\mbox{C}_2\mbox{O}_2$),
was synthesized by Gleizes and Verdaguer and stimulated the public
interest in this potential subject.
The following example of an ordered bimetallic chain \cite{Pei38},
MnCu(pba)(H$_2$O)$_3$$\cdot$$2$H$_2$O
($\mbox{pba}=1,3\mbox{-propylenebis(oxamato)}
 =\mbox{C}_7\mbox{H}_6\mbox{N}_2\mbox{O}_6$),
exhibiting more pronounced one dimensionality, further activated the
physical \cite{Dril13,Verd44}, as well as chemical \cite{Kahn82},
investigations.
There also appeared an idea \cite{Cane56} that the alternating
magnetic centers do not need to be metal ions but may be organic
radicals.

   A practical model of a ferrimagnetic chain is two kinds of spins
$S$ and $s$ ($S>s$) alternating on a ring with antiferromagnetic
exchange coupling between nearest neighbors, as described by the
Hamiltonian,
\begin{equation}
   {\cal H}
      =J\sum_{j=1}^N
        \left(
         \mbox{\boldmath$S$}_{j} \cdot \mbox{\boldmath$s$}_{j}
        +\mbox{\boldmath$s$}_{j} \cdot \mbox{\boldmath$S$}_{j+1}
        \right)\,.
   \label{E:H}
\end{equation}
Here $N$ is the number of unit cells and we set the unit-cell length
equal to $2a$ in the following.
The simplest case, $(S,s)=(1,\frac{1}{2})$, has so far been discussed
fairly well.
There lie both ferromagnetic and antiferromagnetic long-range orders
in the ground state \cite{Tian53,Pati94,Breh21}.
The ground state, which is a multiplet of spin $(S-s)N$, shows
elementary excitations of two distinct types \cite{Yama10}.
The excitations of ferromagnetic aspect, reducing the ground-state
magnetization, form a gapless dispersion relation, whereas those
of antiferromagnetic aspect, enhancing the ground-state
magnetization, are gapped from the ground state.
As a result of the low-energy structure of dual aspect, the specific
heat shows a Shottky-like peak in spite of the ferromagnetic
low-temperature behavior and the magnetic susceptibility times
temperature exhibits a round minimum \cite{Pati94,Yama08,Dril13}.
When the exchange coupling turns anisotropic, the dispersion of the
ferromagnetic excitations is no more quadratic \cite{Alca67} and the
plateau in the ground-state magnetization curve \cite{Kura62} due to
the gapped antiferromagnetic excitations vanishes via the
Kosterlitz-Thouless transition \cite{Saka53}.

   The quantum behavior of the model with higher spins is not yet so
clear as that in the $(S,s)=(1,\frac{1}{2})$ case.
Though Drillon {\it et al.} \cite{Dril13} made the first attempt to
reveal the general behavior of the model, they fixed the
smaller spin to $\frac{1}{2}$ in their argument with particular
emphasis on the problem of the crystal engineering of a
molecule-based ferromagnet$-$the assembly of the highly magnetic
molecular entities within the crystal lattice in a ferromagnetic
fashion.
There also exists an extensive numerical study \cite{Pati94}, but
the leading attention was not necessarily directed to the
consequences of the variation of the constituent spins.
So far the generic behavior, rather than individual features, of
ferrimagnetic mixed-spin chains has been accentuated or predicted,
but there are few attempts to characterize or classify the typical
one-dimensional ferrimagnetic behavior as a function of $(S,s)$.
In such circumstances, we aim in this article at elucidating which
property of the model (\ref{E:H}) is universal and which one, if any,
is variable with $(S,s)$.

   It is true that numerical tools are quite useful in this context,
but they are not almighty.
Although the low-temperature ferromagnetic behavior is quite
interesting from the experimental point of view, it is hardly
feasible to numerically take grand-canonical averages at low enough
temperatures.
It is also unfortunate that with the increase of $(S,s)$, the
available information is more and more reduced in both quality and
quantity.
Then we have an idea of describing the model in terms of the
spin-wave theory.
The conventional spin-wave treatment of low-dimensional magnets may
discourage us.
The Haldane conjecture \cite{Hald64} revealed a qualitative breakdown
of the spin-wave description of one-dimensional Heisenberg
antiferromagnets.
Neither quantum corrections \cite{Oguc17} nor additional constraints
\cite{Reze89} to spin fluctuations end up with an overall scenario
for the low-energy physics applicable to general spins.
However, Ivanov \cite{Ivan24} has recently reported that the
spin-wave description of one-dimensional Heisenberg ferrimagnets is
quite successful.
Though his calculations were restricted to the ground-state energy
and magnetization, the highly accurate estimates obtained there are
surprising enough, considering the diverging ground-state
magnetization in the one-dimensional antiferromagnetic spin-wave
theory.
For antiferromagnets, quantum fluctuations of domain-wall type,
connecting the two degenerate N\'eel states, are important, whereas
for ferrimagnets, domain-wall excitations lead to magnetization
fluctuations and are thus of less significance.
Therefore it is likely that the spin-wave approach is highly
efficient for ferrimagnets.
In an attempt to demonstrate such an idea, we try to describe not
only the low-energy structure but also the thermal behavior of the
Heisenberg ferrimagnetic spin chains (\ref{E:H}) within the framework
of the spin-wave theory.
Extensive numerical calculations, supplemented by the spin-wave
analysis, fully reveal the one-dimensional ferrimagnetic behavior as
a function of the constituent spins.

\section{Elementary Excitations}\label{S:EE}

   In order to investigate the low-energy structure, we employ a new
quantum Monte Carlo technique \cite{Yama09} as well as the
conventional Lanczos diagonalization algorithm.
The idea is in a word expressed as extracting the low-lying
excitations from imaginary-time quantum Monte Carlo data at a low
enough temperature.
The imaginary-time correlation function $S(q,\tau)$ is generally
defined as
\begin{equation}
   S(q,\tau)=\left\langle
   {\rm e}^{{\cal H}\tau}O_q{\rm e}^{-{\cal H}\tau}O_{-q}
             \right\rangle\,,
\end{equation}
where $O_q=N^{-1}\sum_{j=1}^N O_j{\rm e}^{2{\rm i}aqj}$
is the Fourier transform of an arbitrary local operator $O_j$, which
may be an effective combination of the spins $\mbox{\boldmath$S$}$
and $\mbox{\boldmath$s$}$, and the thermal average at a given
temperature $\beta^{-1}=k_{\rm B}T$,
$\left\langle A\right\rangle \equiv
 {\rm Tr}[{\rm e}^{-\beta{\cal H}} A]/
 {\rm Tr}[{\rm e}^{-\beta{\cal H}}]$,
is taken in a certain subspace.
$S(q,\tau)$ can be represented in terms of the eigenvectors and
eigenvalues of the Hamiltonian,
$\vert l;k\rangle$ $(l=1,2,\cdots)$ and
$E_{l}(k)$ ($E_1(k) \leq E_2(k) \leq \cdots$),
and behaves like
\begin{equation}
   S(q,\tau)
     \simeq\sum_{l}
     \left\vert \langle 1;k_0 \vert
     S_q^z
     \vert l;k_0+q \rangle \right\vert^2
     {\rm e}^{-\tau\left[E_{l}(k_0+q)-E_1(k_0)\right]} \,,
\end{equation}
at a sufficiently low temperature, where $k_0$ is the momentum at
which the lowest-energy state in the subspace is located.
Therefore we can reasonably approximate $E_1(k_0+q)-E_1(k_0)$ by the
slope $-\partial{\rm ln}[S(q,\tau)]/\partial\tau$ in the large-$\tau$
region.
When we take interest in the lower edge of the excitation spectrum,
such a treatment is rather straightforward.
The elementary excitations of the Haldane antiferromagnets were
indeed revealed thus \cite{Yama48}.
Here, due to the two kinds of spins in a chain and the resultant dual
aspect of the low-energy structure, the relevant subspace and
operator $O_j$ are not uniquely defined.
Since the total magnetization, $M=\sum_j(S_j^z+s_j^z)$, is a
conserved quantity in the present system, we consider calculating
$S(q,\tau)$ independently in each subspace with a given $M$
\cite{Yama45}.
The elementary excitation energies for the ferromagnetic branch are
obtained from a single calculation, $S(q,\tau)$ in the subspace of
$M=0$, while those for the antiferromagnetic branch from a couple of
calculations, $S(q,\tau)$ and the lowest energy in the subspace of
$M=(S-s)N+1$.
We have taken $S_j^z\pm s_j^z$ for $O_j$ and found that
$O_j=S_j^z-s_j^z$ extracts the eigenvalues of the bonding
(lower-energy) states in both subspaces.
The choice of the scattering matrices is a profound problem in itself
and is fully discussed elsewhere \cite{Yama10,Yama28}.

   Although the chain length we can reach with the
exact-diagonalization method is strongly limited, the diagonalization
results are still helpful in the present system whose correlation
length is generally so small as to be comparable to the unit-cell
length (see Fig. \ref{F:Ss} bellow).
Actually, the ground-state energies for $N=10$ coincide with their
thermodynamic-limit values within the first several digits.
The Lanczos algorithm gives the most precise estimate for the
ground-state energy and the antiferromagnetic excitation gap, whereas
the quantum Monte Carlo technique is necessary for the evaluation of
the curvature of the small-momentum dispersion.

   On the other hand, we consider a spin-wave description of the
elementary excitations as well.
We introduce the bosonic operators for the spin deviation in each
sublattice via
\begin{equation}
   \left.
   \begin{array}{lll}
      S_j^+=\sqrt{2S-a_j^\dagger a_j}\ a_j\,,&
      S_j^z=S-a_j^\dagger a_j\,,\\
      s_j^+=b_j^\dagger\sqrt{2s-b_j^\dagger b_j}\ ,&
      s_j^z=-s+b_j^\dagger b_j\,,
   \end{array}
   \right.
   \label{E:HP}
\end{equation}
where we regard $S$ and $s$ as quantities of the same order.
The Hamiltonian (\ref{E:H}) is expressed in terms of the bosonic
operators as
\begin{equation}
   {\cal H}=E_{\rm class}+{\cal H}_0+{\cal H}_1+O(S^{-1})\,,
   \label{E:Hboson}
\end{equation}
where $E_{\rm class}=-2sSJN$ is the classical ground-state energy,
and ${\cal H}_0$ and ${\cal H}_1$ are the one-body and two-body terms
of the order $O(S^1)$ and $O(S^0)$, respectively.
We may consider the simultaneous diagonalization of ${\cal H}_0$ and
${\cal H}_1$ in the naivest attempt to go beyond the noninteracting
spin-wave theory.
Indeed, the higher-order terms we take into account, the better
description of the ground-state properties we obtain \cite{Ivan24}.
However, such an idea, as a whole, ends in failure, bringing a gap to
the lowest-lying ferromagnetic excitation branch and thus
qualitatively misreading the low-energy physics.
Therefore, we take an alternative approach at the idea of first
diagonalizing ${\cal H}_0$ and next extracting relevant corrections
from ${\cal H}_1$.
${\cal H}_0$ is diagonalized as \cite{Pati94,Breh21}
\begin{equation}
   {\cal H}_0
     =E_0
     +J\sum_k
      \left(
       \omega_{k}^-\alpha_k^\dagger\alpha_k
      +\omega_{k}^+\beta_k^\dagger \beta_k
      \right)\,,
   \label{E:diagH0}
\end{equation}
where
\begin{equation}
   E_0=J\sum_k
   \left[
    \omega_k-(S+s)
   \right]\,,
\end{equation}
is the $O(S^1)$ quantum correction to the ground-state energy, and
$\alpha_k^\dagger$ and $\beta_k^\dagger$ are the creation operators
of the ferromagnetic and antiferromagnetic spin waves of momentum $k$
whose dispersion relations are given by
\begin{equation}
   \omega_{k}^\pm=\omega_k\pm(S-s)\,,
\end{equation}
with
\begin{equation}
   \omega_k=\sqrt{(S-s)^2+4Ss\sin^2(ak)}\,.
\end{equation}
Using the Wick theorem, ${\cal H}_1$ is rewritten as
\begin{eqnarray}
   {\cal H}_1
   &=&E_1-J\sum_k
    \left(
    \delta\omega_k^-\alpha_k^\dagger\alpha_k
    +\delta\omega_k^+\beta_k^\dagger\beta_k
    \right)
   \nonumber\\
   &+&
     {\cal H}_{\rm irrel}+{\cal H}_{\rm quart}\,,
\end{eqnarray}
where
the $O(S^0)$ correction to the ground-state energy, $E_1$, and those
to the dispersions, $\delta\omega_k^\pm$, are, respectively, given by
\widetext
\vskip 4mm

\narrowtext
\begin{flushleft}
\vskip 1mm
\ \ \mbox{\psfig{figure=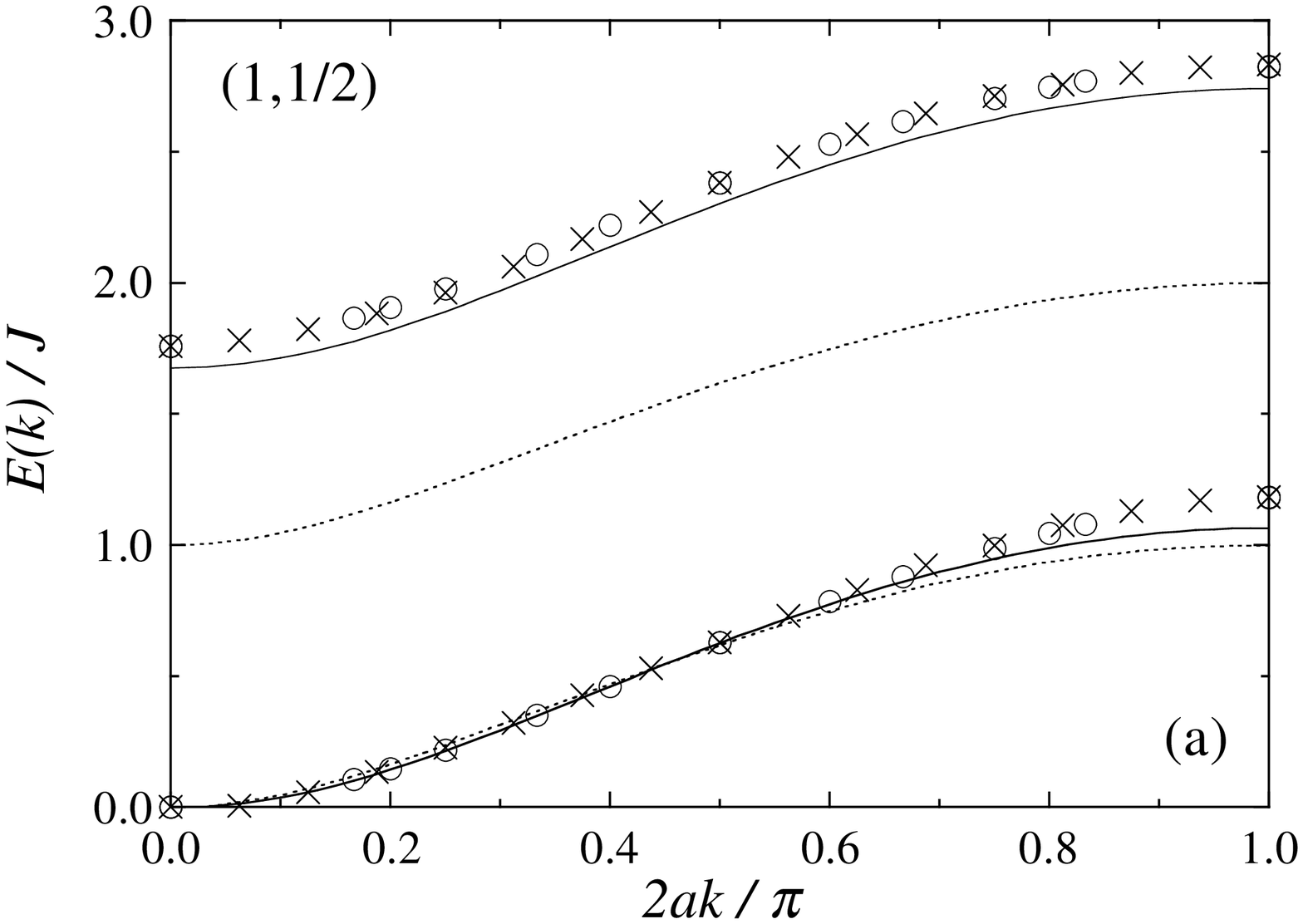,width=80mm,angle=0}}
\vskip -2mm
\ \ \mbox{\psfig{figure=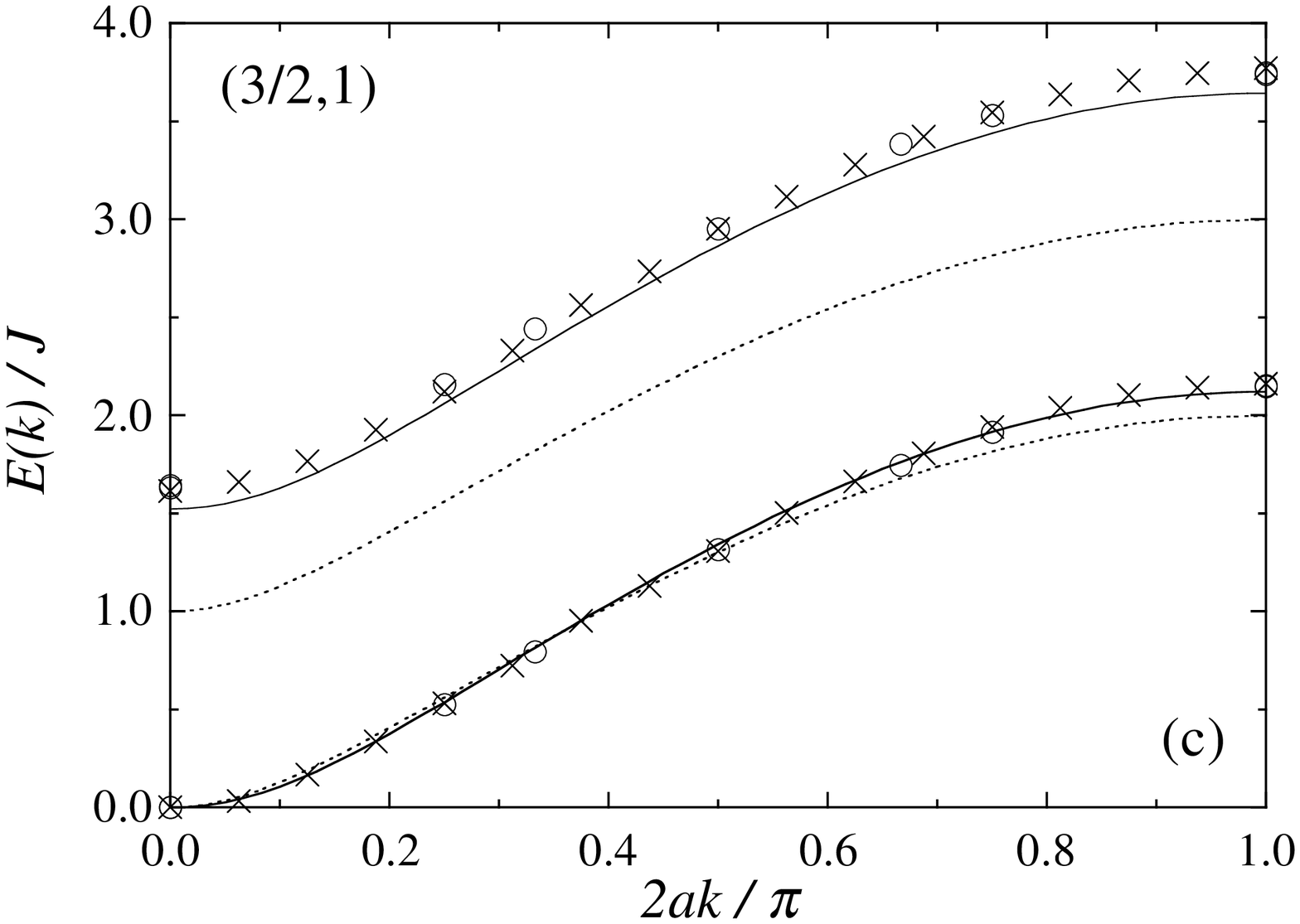,width=80mm,angle=0}}
\vskip 1mm
\ \ \mbox{\psfig{figure=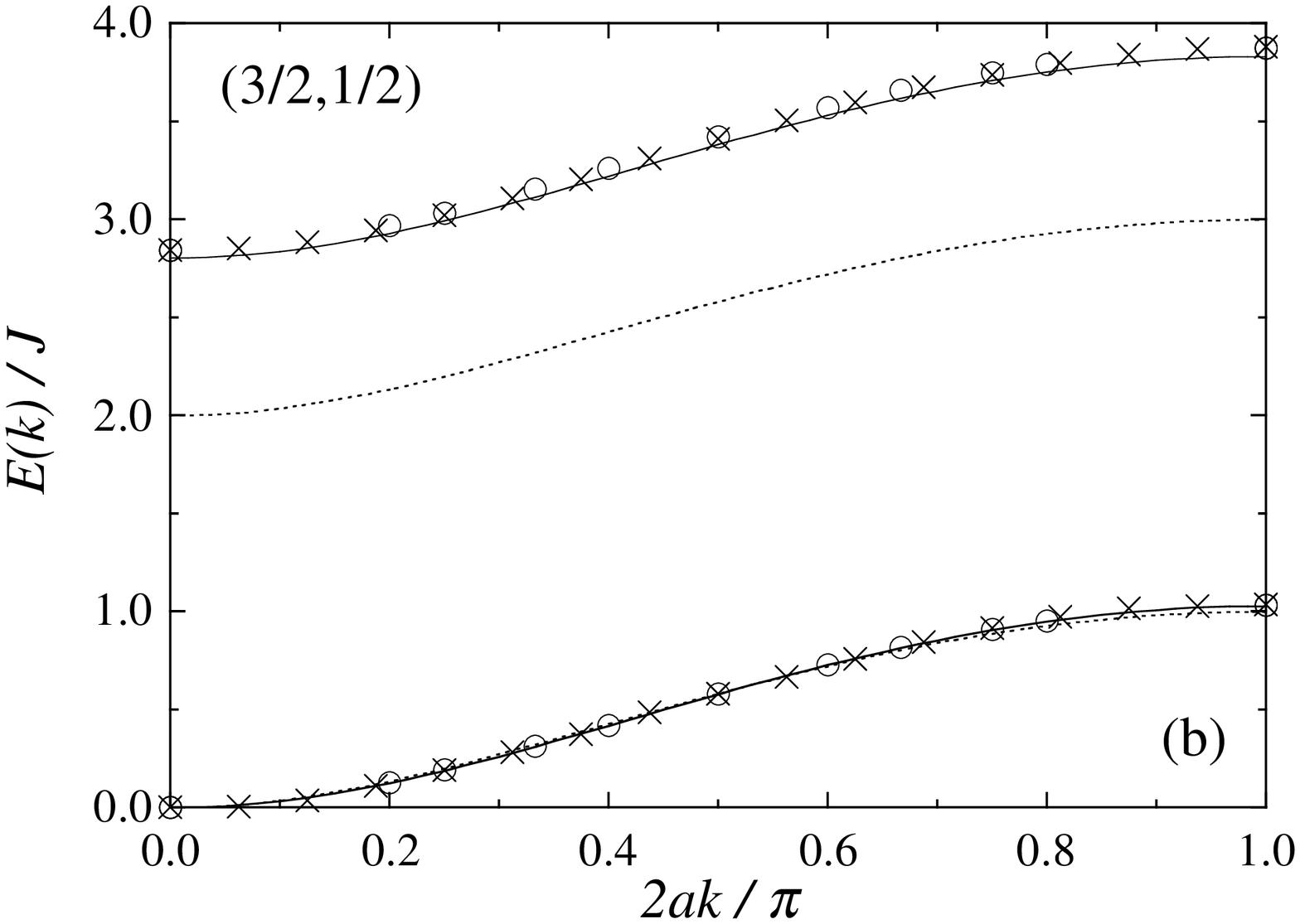,width=80mm,angle=0}}
\vskip -2mm
\ \ \mbox{\psfig{figure=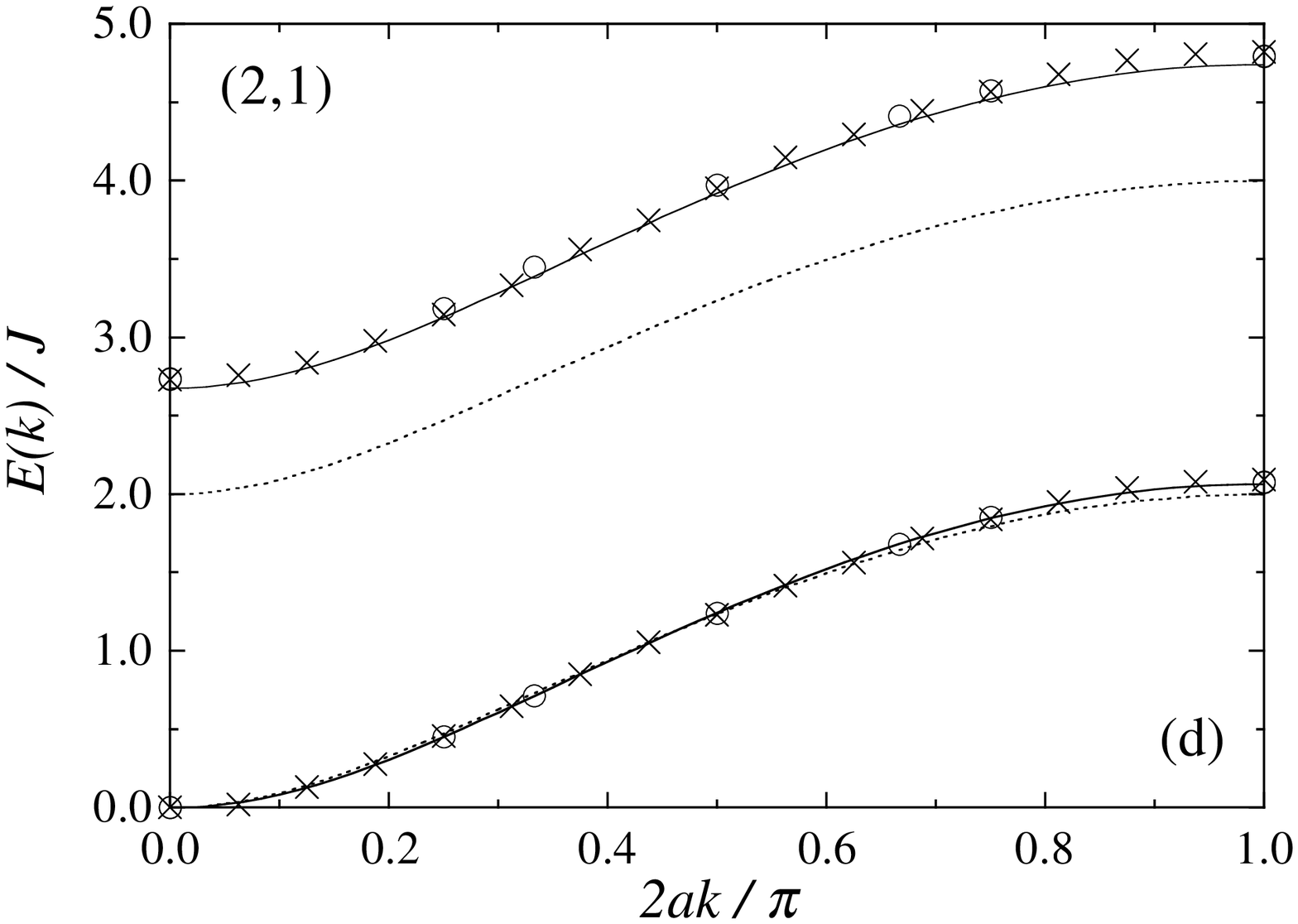,width=80mm,angle=0}}
\end{flushleft}
\widetext
\begin{figure}
\vskip -10mm
\caption{Dispersion relations of the ferromagnetic and
         antiferromagnetic elementary excitations, namely, the
         lowest-energy states in the subspaces of $M=N/2\mp 1$.
         The noninteracting- and interacting-spin-wave calculations
         are shown by dotted and solid lines, respectively, whereas
         $\times$ and $\bigcirc$ represent the quantum Monte Carlo
         estimates ($N=32$) and the exact-diagonalization results
         (up to $N=12$ (a), up to $N=10$ (b), up to $N=8$ (c,$\,$d)),
         respectively.
         Here we plot the excitation energy $E(k)$ taking the
         ground-state energy as zero.}
\label{F:Ek}
\end{figure}
\narrowtext
\begin{eqnarray}
   &&
   E_1=-2JN
       \left[
        {\mit\Gamma}_1^2+{\mit\Gamma}_2^2
       +\left(
         \sqrt{S/s}+\sqrt{s/S}
        \right)
        {\mit\Gamma}_1{\mit\Gamma}_2
       \right]\,,
   \label{E:E1} \\
   &&
   \delta\omega_k^\pm
      =2(S+s){\mit\Gamma}_1
       \frac{\sin^2(ak)}{\omega_k}
      +\frac{{\mit\Gamma}_2}{\sqrt{Ss}}
       \left[
        \omega_k\pm (S-s)
       \right]\,,
    \label{E:domega}
\end{eqnarray}
with
\begin{equation}
   \left.
   \begin{array}{l}
     {\mit\Gamma}_1
     ={\displaystyle\frac{1}{2N}\sum_k}
      \left(
       {\displaystyle\frac{S+s}{\omega_k}-1}
      \right)\,,\\
     {\mit\Gamma}_2
     =-{\displaystyle\frac{1}{N}\sum_k}
       {\displaystyle\frac{\sqrt{Ss}\,{\rm cos}^2(ak)}
                          {\omega_k}}\,,\\
   \end{array}
   \right.
   \label{E:Gamma}
\end{equation}
while the irrelevant one-body terms
\begin{equation}
   {\cal H}_{\rm irrel}
   =-J\frac{(S-s)^2}{\sqrt{Ss}}{\mit\Gamma}_1
    \sum_k
    \frac{{\rm cos}(ak)}{\omega_k}
    (\alpha_k\beta_k+\alpha_k^\dagger\beta_k^\dagger)\,,
\end{equation}
and the residual two-body interactions ${\cal H}_{\rm quart}$ are
both neglected so as to keep the ferromagnetic branch gapless.
The resultant Hamiltonian is compactly represented as
\begin{equation}
   {\cal H}
     \simeq E_{\rm g}
    +J\sum_k
     \left(
      {\widetilde\omega}_k^- \alpha_k^\dagger \alpha_k
     +{\widetilde\omega}_k^+ \beta_k^\dagger  \beta_k
     \right)\,,
\end{equation}
with
\begin{eqnarray}
   &&
   {\widetilde\omega}_k^\pm=\omega_k^\pm-\delta\omega_k^\pm\,,
   \\
   &&
   E_{\rm g}=E_{\rm class}+E_0+E_1\,.
\end{eqnarray}

   Now we compare all the calculations in Fig. \ref{F:Ek}.
The diagonalization results fully demonstrate the steady
applicability and good precision of our Monte Carlo treatment.
The spin-wave approach generally gives a good description of the
low-energy structure.
Even the free spin waves allow us to have a qualitative view of the
elementary excitations.
The relatively poor description of the antiferromagnetic branch by
the free spin waves implies that the quantum effect is more relevant
in the antiferromagnetic branch.
Here we may be reminded of the spin-wave treatment of mono-spin
Heisenberg magnets.
For the ferromagnetic chains, the spin-wave dispersions are nothing
but the exact picture of the elementary excitations, while for the
antiferromagnetic chains, those are generally no more than a
qualitative view.
The point is that in the present system the spin-wave picture is
efficient for both elementary excitation branches and the spin-wave
series potentially lead to an accurate description.
More specifically, the divergence of the boson numbers, which plagues
the antiferromagnetic spin-wave treatment in one dimension, does not
occur in the present system.
This viewpoint is further discussed in the final section.
We conclude this section by listing in Table \ref{T:comp} the
spin-wave estimates of a few interesting quantities in comparison
with the numerical solutions, where we define the curvature $v$ as
${\widetilde\omega}_{k\rightarrow 0}^-=v(2ak)^2$.

\section{Thermodynamic Properties}\label{S:ThP}

   The dual structure of the excitations leads to unique thermal
behavior.
In order to complement quantum Monte Carlo thermal calculations,
especially at low temperatures, we consider describing the
thermodynamics in terms of the spin-wave theory.
Introducing the additional constraint of the total magnetization
being zero into the conventional spin-wave theory, Takahashi
\cite{Taka68} succeeded in obtaining an excellent description of the
low-temperature thermal behavior of one-dimensional Heisenberg
ferromagnets.
The present authors have recently applied the idea to the
spin-$(1,\frac{1}{2})$ ferrimagnetic Heisenberg chain \cite{Yama08}.
Here we develop the method for general spin cases and make a
detailed analysis of its validity as a function of $(S,s)$.
The core idea of the so-called modified spin-wave theory
\cite{Taka94,Hirs69} can be summarized as controlling the boson
numbers by imposing a certain constraint on the magnetization.
From this point of view, the zero-magnetization constraint, which
is quite reasonable for isotropic magnets, plays a relevant role in
ferromagnets.
The resultant low-temperature expansions \cite{Taka68} of the
specific heat and susceptibility of the spin-$s$ Heisenberg
ferromagnetic chain,
\begin{eqnarray}
   &&
   \frac{C}{Nk_{\rm B}}
    =\frac{3}{8s^{\frac{1}{2}}}
     \frac{\zeta(\frac{3}{2})}{\sqrt{\pi}}
     t^{\frac{1}{2}}
    -\frac{1}{2s^2}t
    +O(t^{\frac{3}{2}})
   \,,\label{E:MSWferroC}\\
   &&
   \frac{\chi J}{N(g\mu_{\rm B})^2}
    =\frac{2s^4}{3}t^{-2}
    -s^{\frac{5}{2}}\frac{\zeta(\frac{1}{2})}{\sqrt{\pi}}
     t^{-\frac{3}{2}}
   \nonumber\\
   && \qquad\qquad\quad\ \ 
    +\frac{s}{2}
     \left[
      \frac{\zeta(\frac{1}{2})}{\sqrt{\pi}}
     \right]^2 t^{-1}
    +O(t^{-\frac{1}{2}})
   \,,\label{E:MSWferroS}
\end{eqnarray}
with $t=k_{\rm B}T/J$ and Riemann's zeta function $\zeta(z)$, indeed
coincide with the thermodynamic Bethe-ansatz calculations
\cite{Taka08} for $s=\frac{1}{2}$ within the leading few terms.

   In the present system, the zero-magnetization constraint is
explicitly represented as 
\begin{equation}
   \sum_j\langle S_j^z+s_j^z\rangle
     =(S-s)N
     -\sum_k\sum_{\sigma=\pm}
      \sigma{\widetilde n}_k^{-\sigma}
     =0\,,
   \label{E:M}
\end{equation}
where $n_k^\pm=\sum_{n^-,n^+}n^\pm P_k(n^-,n^+)$ with $P_k(n^-,n^+)$
being the  probability of $n^-$ ferromagnetic and $n^+$
antiferromagnetic spin waves appearing in the $k$-momentum state.
Equation (\ref{E:M}) straightforwardly proposes that the thermal spin
deviation should cancel the N\'eel-state magnetization.
By minimizing the free energy
\begin{eqnarray}
   &&
   F=E_{\rm g}
    +\sum_k
     (\widetilde n^-_k\widetilde\omega_k^-
     +\widetilde n^+_k\widetilde\omega_k^+)
   \nonumber\\
   && \qquad\,
    +k_{\rm B}T
     \sum_k\sum_{n^-,n^+}P_k(n^-,n^+){\rm ln}P_k(n^-,n^+)\,,
   \label{E:F}
\end{eqnarray}
with respect to $P_k(n^-,n^+)$ at each $k$ under the condition
(\ref{E:M}) as well as the trivial constraints
$\sum_{n^-,n^+}P_k(n^-,n^+)=1$, we obtain
\begin{eqnarray}
   &&
   \frac{C}{Nk_{\rm B}}
     =\frac{3}{4}\left(\frac{S-s}{Ss}\right)^{\frac{1}{2}}
      \frac{\zeta(\frac{3}{2})}{\sqrt{2\pi}}
      \widetilde t^{\frac{1}{2}}
     -\frac{1}{Ss}\widetilde t 
     +O(\widetilde t^{\frac{3}{2}})\,,
   \label{E:MSWferriC}\\
   &&
   \frac{\chi J}{N(g\mu_{\rm B})^2}
     =\frac{Ss(S-s)^2}{3}\widetilde t^{-2}
     -(Ss)^{\frac{1}{2}}(S-s)^{\frac{3}{2}}
      \frac{\zeta(\frac{1}{2})}{\sqrt{2\pi}}
      \widetilde t^{-\frac{3}{2}}
   \nonumber \\
   && \qquad\qquad\quad\ \ 
     +(S-s)\left[
            \frac{\zeta(\frac{1}{2})}{\sqrt{2\pi}}
           \right]^2 \widetilde t^{-1}
     +O(\widetilde t^{-\frac{1}{2}})\,,
   \label{E:MSWferriS}
\end{eqnarray}
where $\widetilde t=k_{\rm B}T/J\gamma$ with
$\gamma=1-{\mit\Gamma}_1(S+s)/Ss-{\mit\Gamma}_2/\sqrt{Ss}$.
The specific heat $C$ has been obtained by differentiating the free
energy $F$, whereas the susceptibility $\chi$ by calculating
$(\langle M^2\rangle-\langle M\rangle^2)/3T$, where we have set the
$g$ factors of the spins $\mbox{\boldmath$S$}$ and
$\mbox{\boldmath$s$}$ both equal to $g$, because the difference
between them amounts to at most several percent of themselves in
practice \cite{Hagi09}.
\vskip 4mm
\qquad\mbox{\psfig{figure=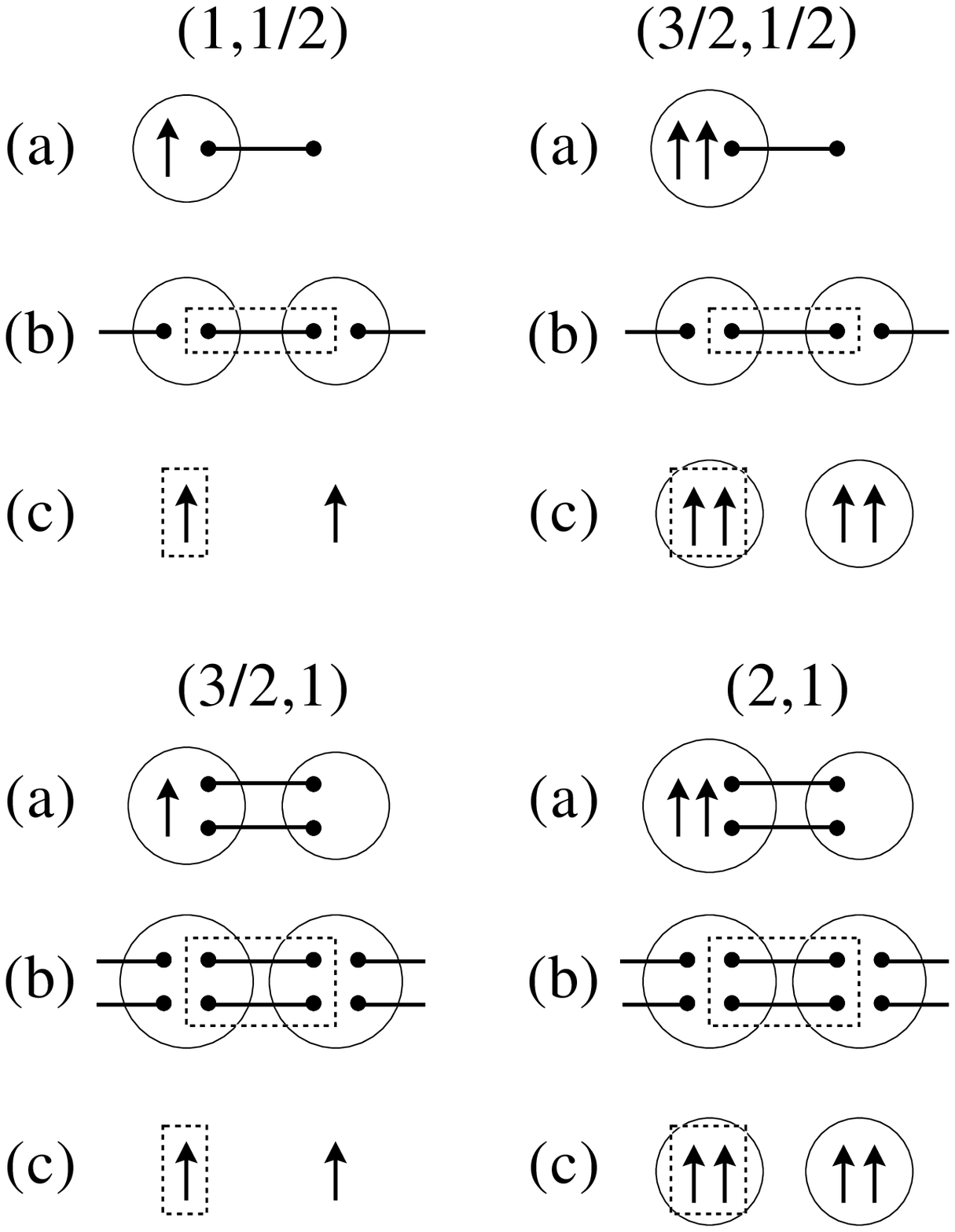,width=80mm,angle=0}}
\begin{figure}
\vskip 0mm
\caption{Schematic representations of the $M=(S-s)N$ ground states of
         spin-$(S,s)$ ferrimagnetic chains of $N$ elementary cells in
         the decoupled-dimer limit (a), the AKLT ground states of
         spin-$(2s)$ antiferromagnetic chains of $2N$ spins (b), the
         $M=2N(S-s)$ ground states of spin-$(S-s)$ ferromagnetic
         chains of $2N$ spins (c).
         The arrow (the bullet symbol) denotes a spin $1/2$ with its
         fixed (unfixed) projection value.
         The solid segment is a singlet pair.
         The circle represents an operation of constructing spins $S$
         and $s$ by symmetrizing the spin $1/2$'s inside.
         The relation ${\rm(a)}\approx[{\rm(b)}+{\rm(c)}]/2$ may be
         expected.}
\label{F:illust}
\end{figure}
\vskip 2mm

   The analytic expressions (\ref{E:MSWferriC}) and
(\ref{E:MSWferriS}) give us a bird's-eye view of the one-dimensional
ferrimagnetic behavior.
The spin-$(S,s)$ ferrimagnet turns into the spin-$s$ antiferromagnet
in the limit of $S\rightarrow s$, whereas it looks like the spin-$S$
ferromagnet in the limit of $S/s\rightarrow\infty$.
In this sense, the subtraction $S-s$ may be regarded as the
ferromagnetic contribution, while the residual spin amplitude $2s$
as the antiferromagnetic one.
No ferromagnetic aspect at $S/s=1$, while a hundred percent
ferromagnetic aspect for $S/s\rightarrow\infty$.
Another consideration also leads us to such a picture.
Since the perturvation from the decoupled dimers \cite{Yama10}
qualitatively well describes the low-lying excitations of the system,
we propose in Fig. \ref{F:illust} an idea of decomposing ferrimagnets
into ferromagnets and antiferromagnets, where we let the decoupled
dimers and the Affleck-Kennedy-Lieb-Tasaki valence-bond-solid states
\cite{Affl99} symbolize ferrimagnets and integer-spin gapped
antiferromagnets, respectively.
Now we expect spin-$(S,s)$ ferrimagnets to behave like combinations
of spin-$(2s)$ antiferromagnets and spin-$(S-s)$ ferromagnets.
Since the antiferromagnetic excitations of the ferrimagnetic ground
state are gapped, the low-temperature behavior of ferrimagnets should
be only of ferromagnetic aspect.
At low temperatures there is indeed no effective contribution from
the spin-$(2s)$ antiferromagnetic chain with an excitation gap
\cite{Hald64} immediately above the ground state.
In this context, we are surprised but pleased to find that provided
$S=2s$, the expressions (\ref{E:MSWferriC}) and (\ref{E:MSWferriS})
coincide with those for ferromagnets, (\ref{E:MSWferroC}) and
(\ref{E:MSWferroS}), except for the quantum renormalizing factor
$\gamma$.
The low-temperature thermodynamics should be dominated by the
small-momentum ferromagnetic excitations.
Within the linearized spin-wave theory, the small-momentum
dispersions of Heisenberg ferrimagnets and ferromagnets are
characterized by the curvatures
\begin{eqnarray}
   &&
   v^{(S,s){\scriptstyle\mbox{-}}{\rm ferri}}
     =\frac{Ss}{2(S-s)}J\,,
   \label{E:vferri}\\
   &&
   v^{(S-s){\scriptstyle\mbox{-}}{\rm ferro}}
     =(S-s)J\,,
   \label{E:vferro}
\end{eqnarray}
respectively, where we find that they coincide with each other only
when $S=2s$.
The criterion, $S=2s$, is further convincing when we consider the
high-temperature behavior.
The paramagnetic behavior of the spin-$(S,s)$ ferrimagnet is given as
\begin{eqnarray}
   &&
   \frac{F^{(S,s){\scriptstyle\mbox{-}}{\rm ferri}}}{Nk_{\rm B}T}
   =-{\rm ln} \left[(2S+1)(2s+1)\right]\,,
   \label{E:HTferriF}\\
   &&
   \frac{k_{\rm B}T\chi^{(S,s){\scriptstyle\mbox{-}}{\rm ferri}}}
        {Ng^2\mu_{\rm B}^2}
   =\frac{1}{3} \left[S(S+1)+s(s+1)\right]\,,
   \label{E:HTferriS}
\end{eqnarray}
whereas those of the spin-$(S-s)$ ferromagnet and the spin-$(2s)$
antiferromagnet are as follows:
\begin{eqnarray}
   &&
   \frac{F^{(S-s){\scriptstyle\mbox{-}}{\rm ferro}}
        +F^{ (2s){\scriptstyle\mbox{-}}{\rm antiferro}}}
        {Nk_{\rm B}T}
   \nonumber\\
   && \qquad
   =-{\rm ln} \left[(2S-2s+1)(4s+1)\right]\,,
   \label{E:HTcombF}\\
   &&
   \frac{k_{\rm B}T
         (\chi^{(S-s){\scriptstyle\mbox{-}}{\rm ferro}}
         +\chi^{ (2s){\scriptstyle\mbox{-}}{\rm antiferro}})}
        {Ng^2\mu_{\rm B}^2}
   \nonumber\\
   && \qquad
   =\frac{1}{3} \left[(S-s)(S-s+1)+2s(2s+1)\right]\,.
   \label{E:HTcombS}
\end{eqnarray}
These asymptotic values agree with each other only when $S=2s$.
This is simply the consequence of the degrees of freedom.
In ferrimagnets of $S>2s$, the ferromagnetic spin degrees of freedom
overbalance the antiferromagnetic ones, while in ferrimagnets of
$S<2s$, vice versa.
Only the {\it balanced} ferrimagnet with $S=2s$ is well approximated
as the simple combination of the spin-$(S-s)$ ferromagnet and the
spin-$(2s)$ antiferromagnet.
However, we note that even in the case of $S=2s$, the similarity
between the ferrimagnetic behavior, (\ref{E:MSWferriC}) and
(\ref{E:MSWferriS}), and the ferromagnetic one,
(\ref{E:MSWferroC}) and (\ref{E:MSWferroS}), does not go beyond the
leading few terms shown here.
The ferromagnetic features of ferrimagnets are thermally smeared out.
On the other hand, the ferrimagnetic behavior further deviates from
the purely ferromagnetic one due to the quantum effect characterized
by $\gamma$.
The low-temperature expansions (\ref{E:MSWferriC}) and
(\ref{E:MSWferriS}) imply that as temperature goes to zero, the
quantum effect is reduced for the specific heat $C$, whereas enhanced
for the susceptibility $\chi$.
In the limit of $S/s\rightarrow\infty$, the quantum corrections
${\mit\Gamma}_1$ and ${\mit\Gamma}_2$ both vanish.

   Although the spin-wave theory combined with the additional
constraint (\ref{E:M}) is so useful, it should further be modified at
higher temperatures so as to control the total number of the bosons
$\sum_{\sigma=\pm}{\widetilde n}_k^{\sigma}$ rather than the
subtraction $\sum_{\sigma=\pm}\sigma{\widetilde n}_k^{-\sigma}$.
In our naivest attempt to improve the theory, we replace the
constraint (\ref{E:M}) by
\begin{eqnarray}
   &&
   \sum_j\langle S_j^z-s_j^z\rangle
      =(S+s-2{\mit\Gamma}_1)N
   \nonumber\\
   && \qquad\qquad\qquad\ \ 
      -(S+s)\sum_k\sum_{\sigma=\pm}
       \frac{\widetilde n^\sigma_k}{\omega_k}
      =0\,.
   \label{E:Mst}
\end{eqnarray}
However, the alternative condition (\ref{E:Mst}) changes the
low-temperature description, (\ref{E:MSWferriC}) and
(\ref{E:MSWferriS}), which should be kept unchanged under any
artificial constraint, as well as considerably underestimates the
Schottky-like characteristic peak of the specific heat.
In an attempt to remove ${\mit\Gamma}_1$ from Eq. (\ref{E:Mst}), we
reach a phenomenological modification
\begin{equation}
   \sum_j\langle :S_j^z-s_j^z:\rangle
      =(S+s)N
      -(S+s)\sum_k\sum_{\sigma=\pm}
       \frac{\widetilde n^\sigma_k}{\omega_k}
      =0\,,
   \label{E:Mst:}
\end{equation}

\widetext
\quad
\vskip 2mm

\narrowtext
\begin{flushleft}
\vskip 0mm
\ \ \mbox{\psfig{figure=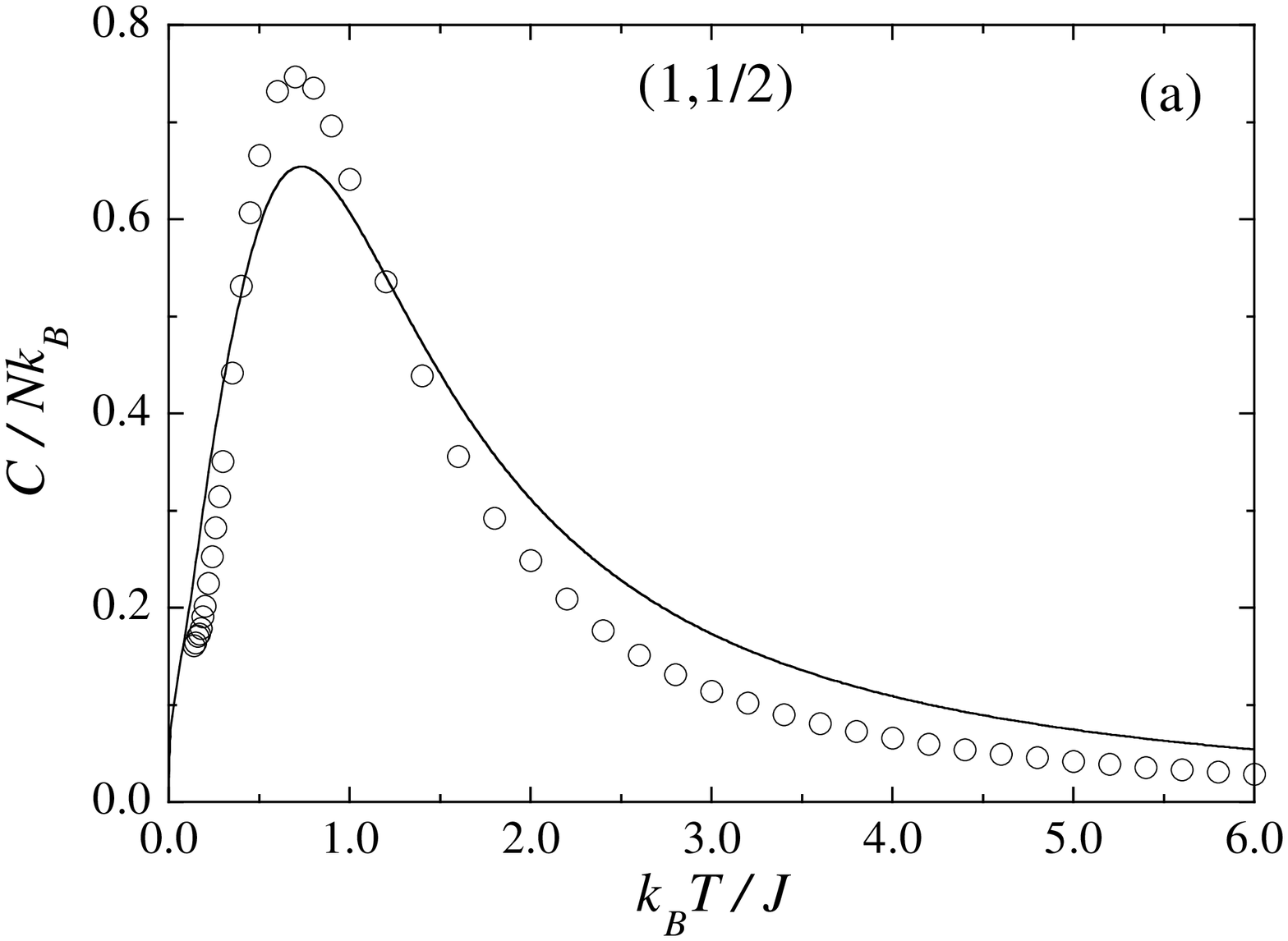,width=80mm,angle=0}}
\vskip -2mm
\ \ \mbox{\psfig{figure=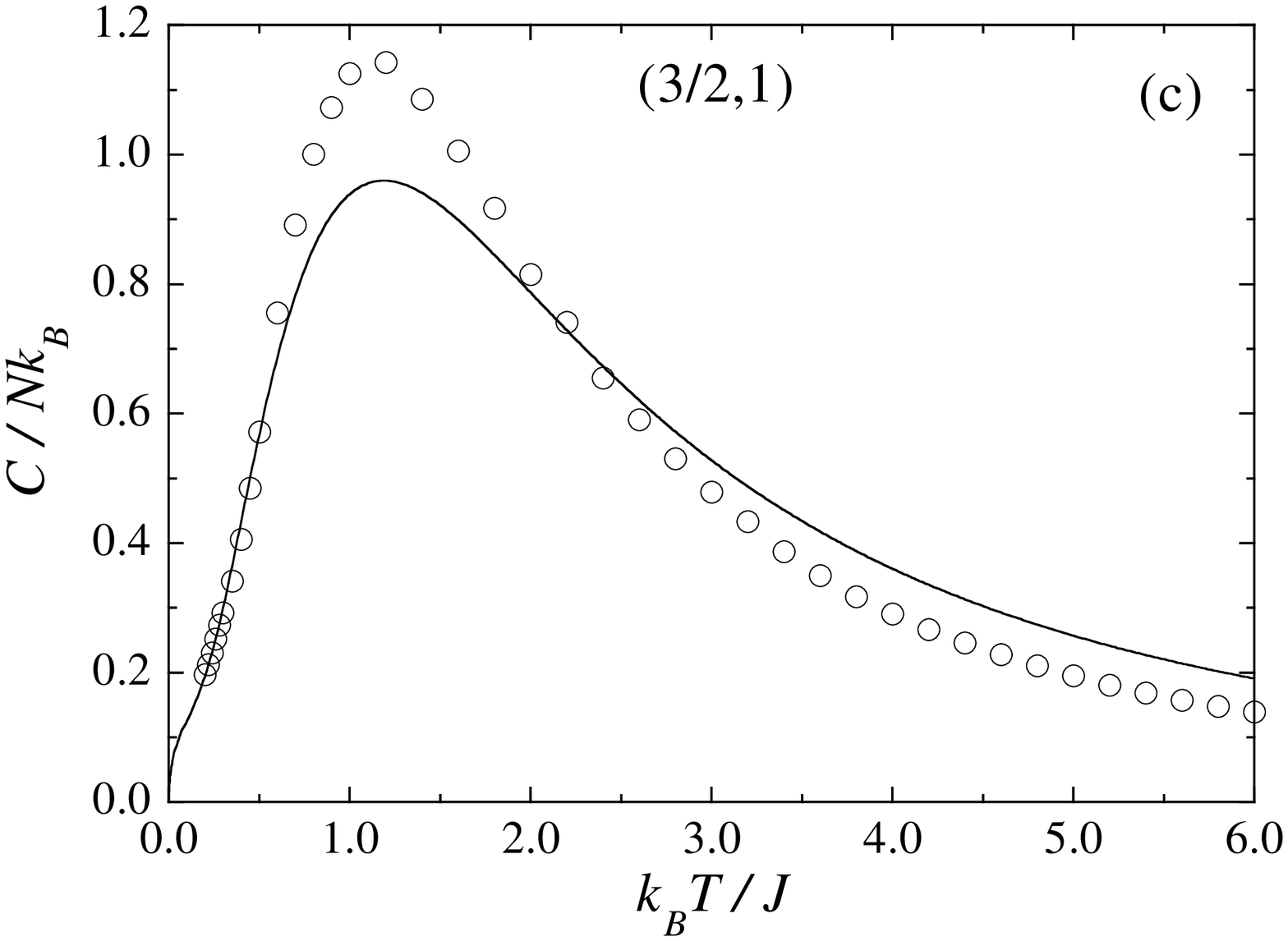,width=80mm,angle=0}}
\vskip 0mm
\ \ \mbox{\psfig{figure=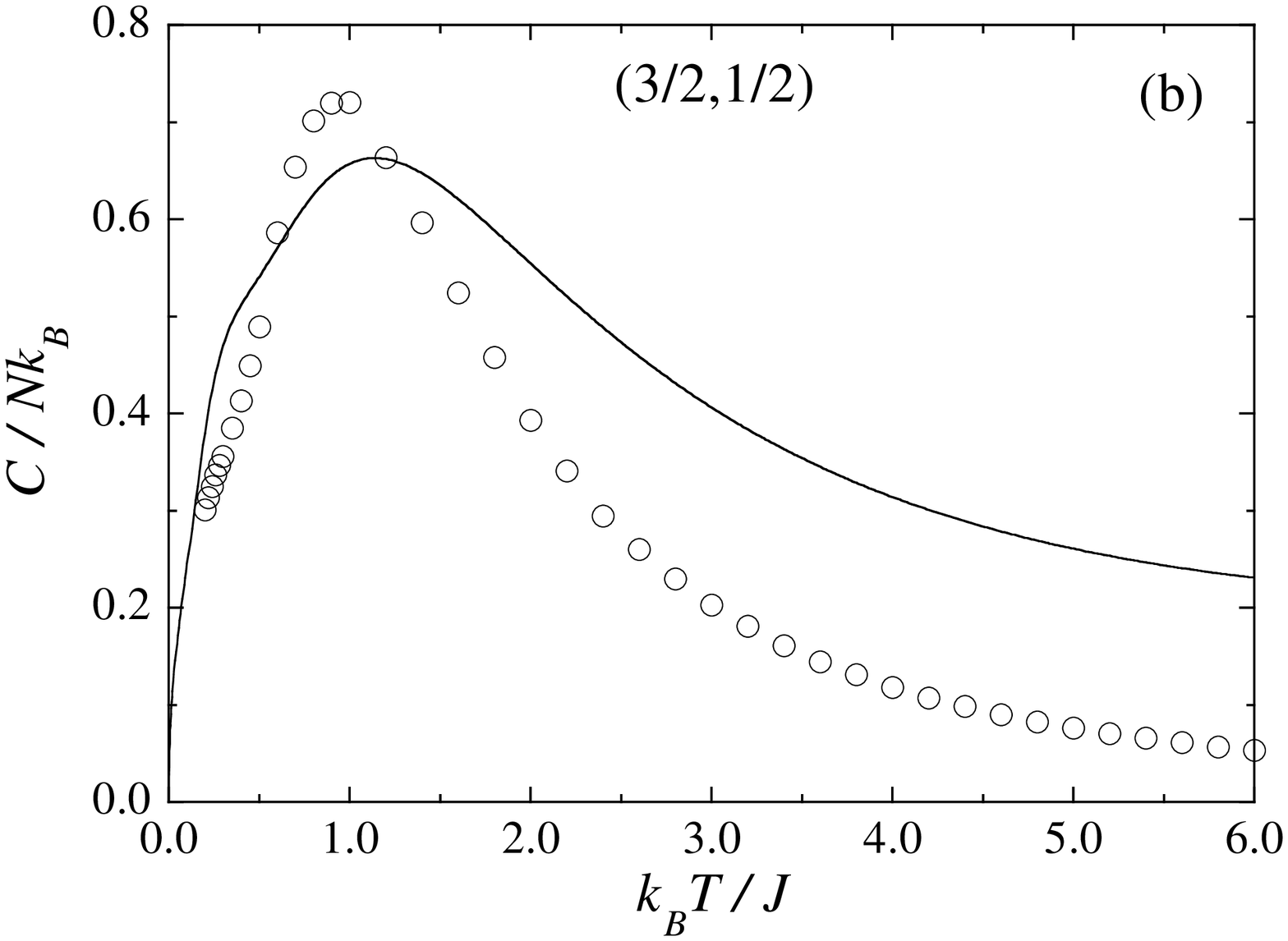,width=80mm,angle=0}}
\vskip -2mm
\ \ \mbox{\psfig{figure=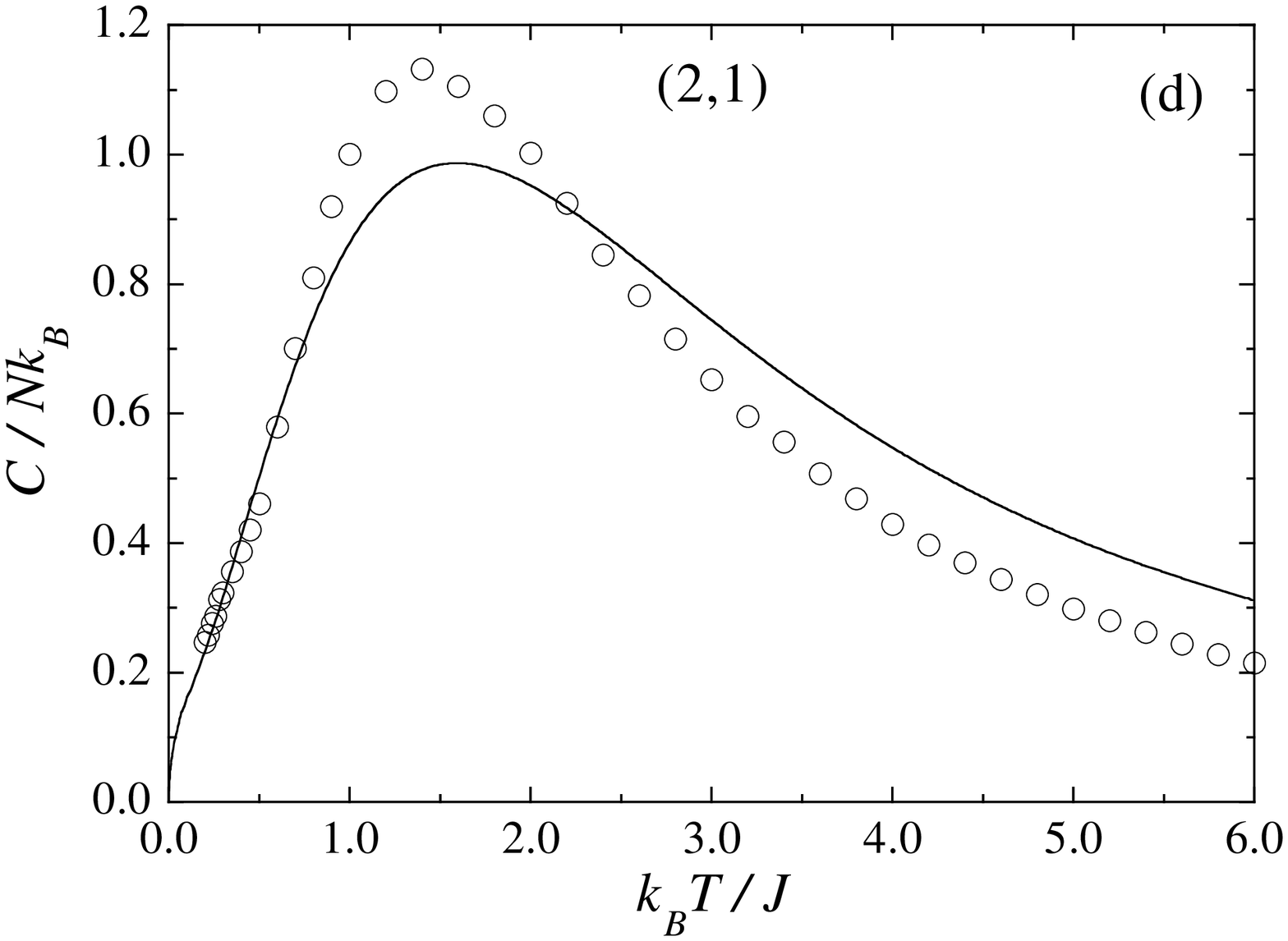,width=80mm,angle=0}}
\end{flushleft}
\widetext
\begin{figure}
\vskip -8mm
\caption{Temperature dependences of the specific heat.
         $\bigcirc$ represents the quantum Monte Carlo estimates,
         whereas the solid line shows the modified-spin-wave
         calculations starting from the interacting-spin-wave
         dispersion relations.}
\label{F:C}
\end{figure}
\narrowtext

\begin{flushleft}
\vskip 0mm
\ \ \mbox{\psfig{figure=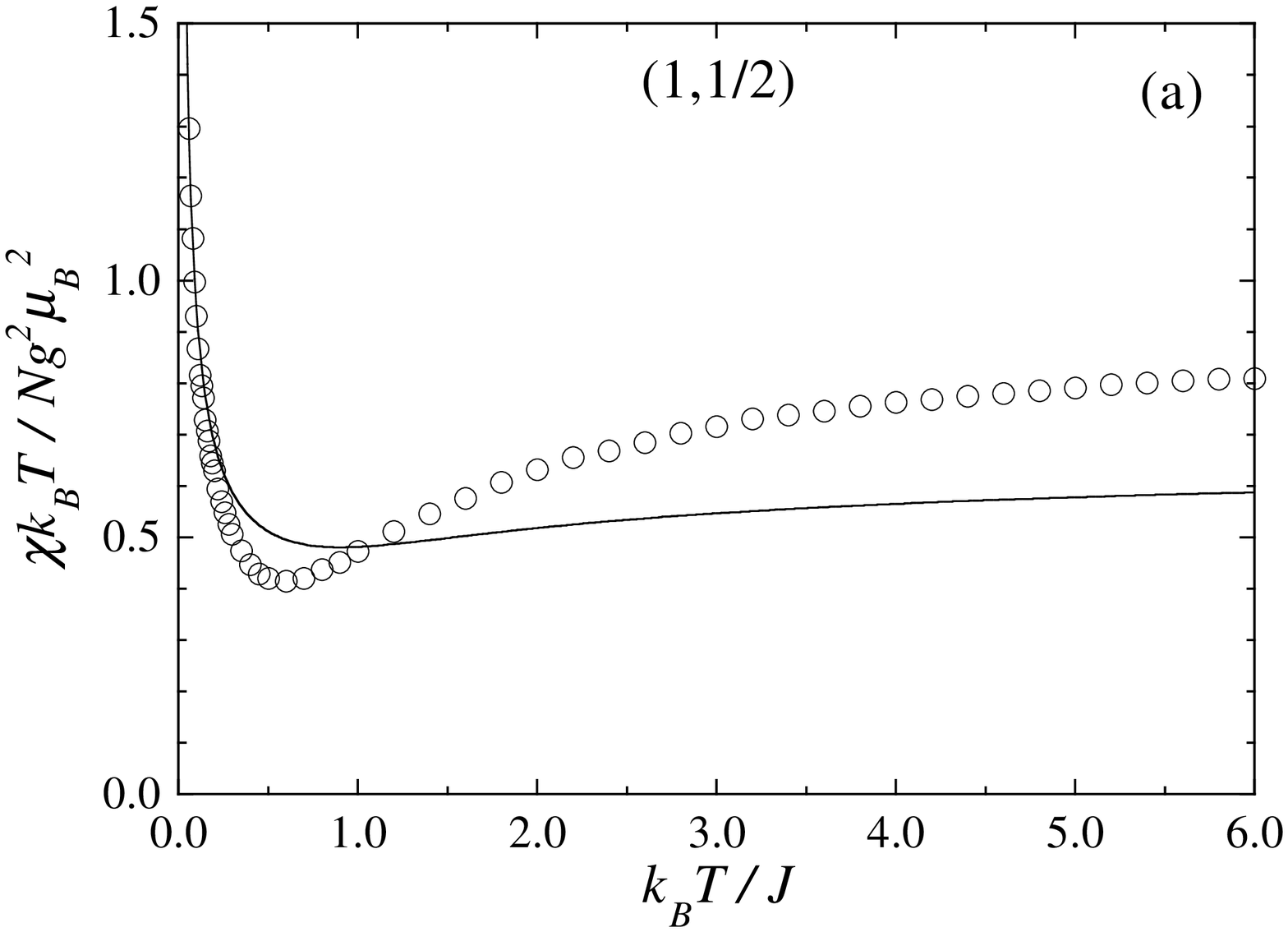,width=80mm,angle=0}}
\vskip -2mm
\ \ \mbox{\psfig{figure=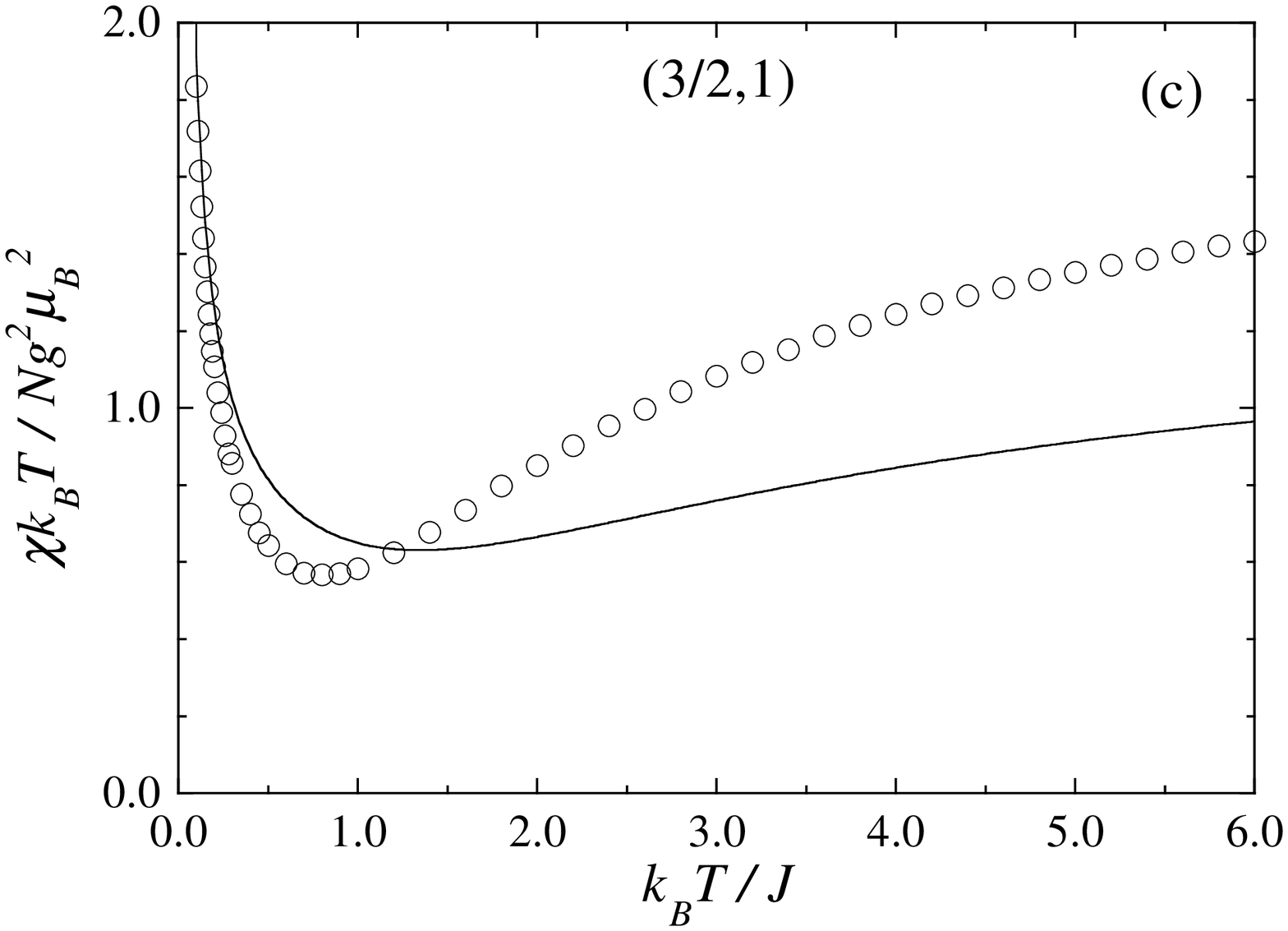,width=80mm,angle=0}}
\vskip 0mm
\ \ \mbox{\psfig{figure=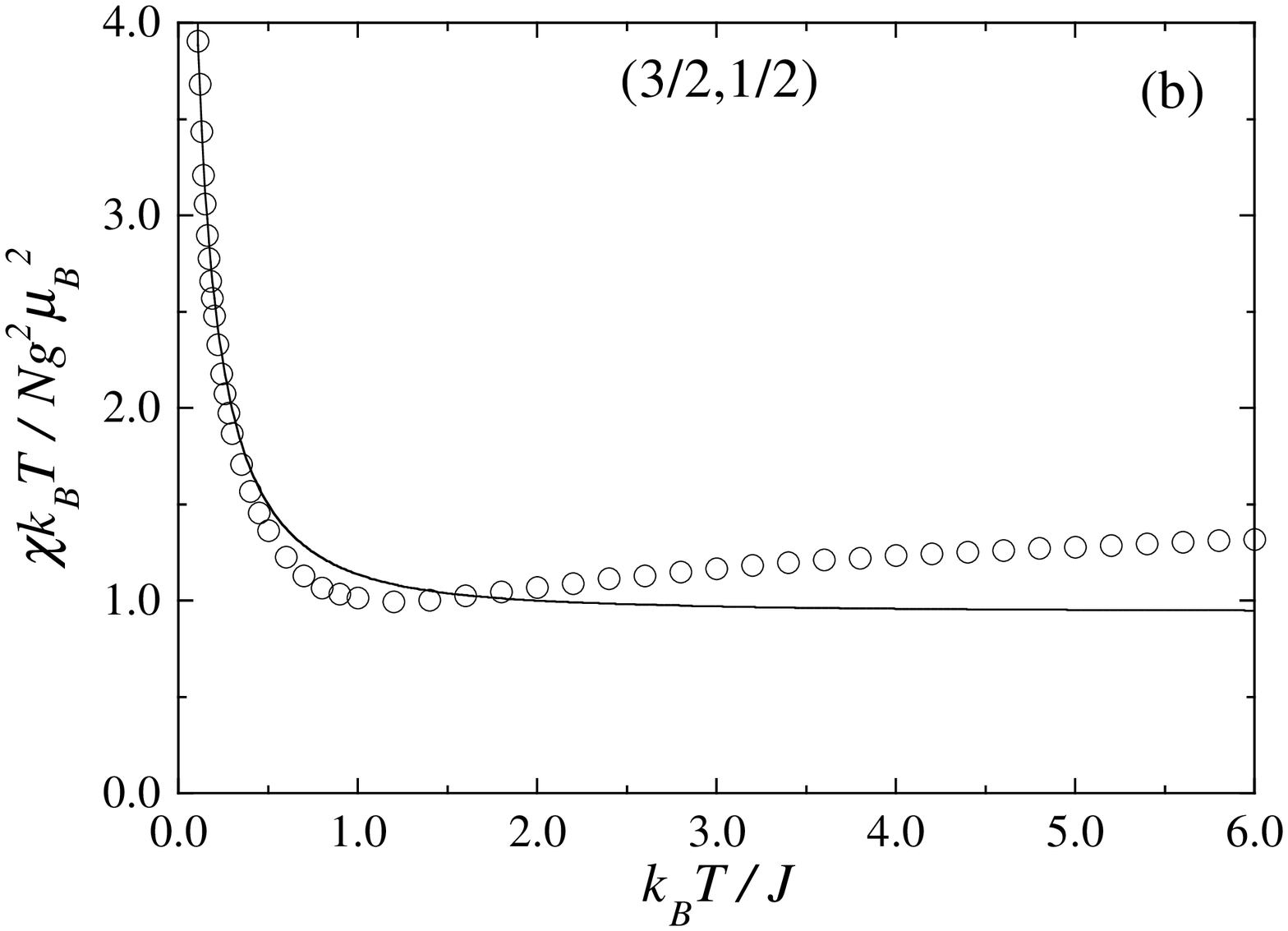,width=80mm,angle=0}}
\vskip -2mm
\ \ \mbox{\psfig{figure=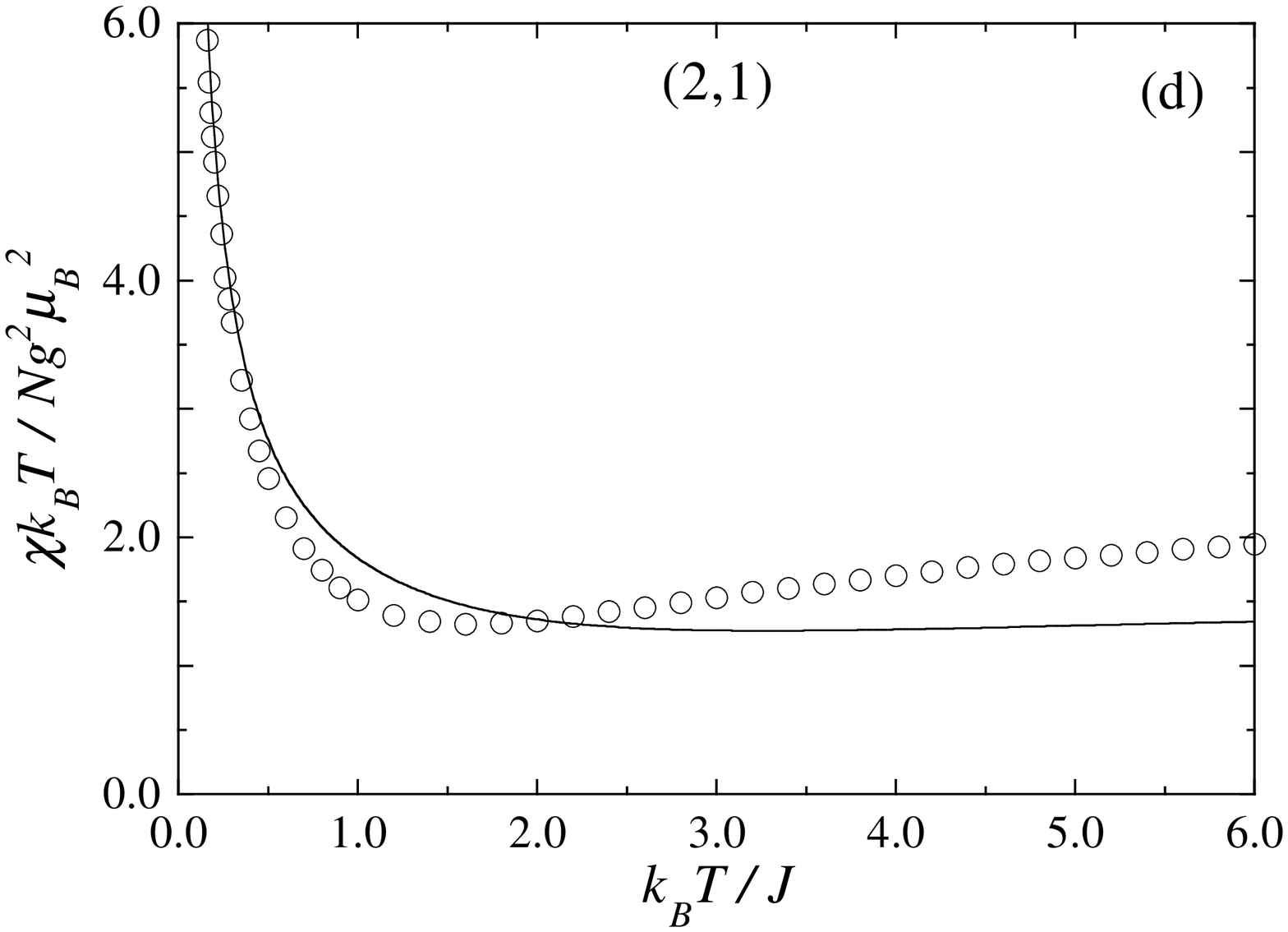,width=80mm,angle=0}}
\end{flushleft}
\widetext
\begin{figure}
\vskip -8mm
\caption{Temperature dependences of the magnetic susceptibility times
         temperature.
         $\bigcirc$ represents the quantum Monte Carlo estimates,
         whereas the solid line shows the modified-spin-wave
         calculations starting from the interacting-spin-wave
         dispersion relations.}
\label{F:chiT}
\end{figure}
\narrowtext
which proposes that the thermal fluctuation should cancel the
N\'eel-state magnetization and results in the same low-temperature
description as Eqs. (\ref{E:MSWferriC}) and (\ref{E:MSWferriS}),
where the normal ordering is taken with respect to both operators
$\alpha$ and $\beta$.
Now the overall description of the thermal quantities is numerically
obtained and is shown, together with the precise quantum Monte Carlo
calculations, in Figs. \ref{F:C} and \ref{F:chiT}.
We stress that these are the spin-wave description of
{\it one-dimensional} magnets.
We are allowed to recognize that the spin-wave picture is
qualitatively valid even at high temperatures.
The spin-wave theory, which is to overestimate the spin degrees of
freedom, inevitably overestimates the energy-derivative quantities
and under estimate the magnetization-derivative ones at high
temperatures.
However, the location of the Schottky-like peak of the specific heat
is generally reproduced well, which should be attributed to the fine
description of the antiferromagnetic excitation gap by the
interacting spin waves.
One might say that the spin-wave description is a overlikely portrait
of the actual behavior.
For the {\it ferromagnetic} ferrimagnet of
$(S,s)=(\frac{3}{2},\frac{1}{2})$, for instance, the low-temperature
shoulder of $C$ is too pronounced and the high-temperature
antiferromagnetic increase of $\chi T$ is too suppressed in the
spin-wave theory.
We note finally in this section that the double constraint does not
improve the theory at all.
At high temperatures, the constraint (\ref{E:Mst:}) dominates the
thermal behavior but the constraint (\ref{E:M}) works little.
At low temperatures, where
${\widetilde n}_k^{-}\gg{\widetilde n}_k^{+}$, the two constraints
(\ref{E:M}) and (\ref{E:Mst:}) are almost degenerate and no numerical
solution is found for the couple of Lagrange multipliers due to the
double constraint.
Further inclusion of any tricky constraint is likely to make us lose
sight of the physical basis.

\section{Discussion}\label{S:D}

   We have studied the low-energy structure and the thermal
properties of Heisenberg ferrimagnetic spin chains featuring the
spin-wave theory.
Even with the modified spin-wave theory, it is still hard to describe
the thermodynamics over the whole temperature range.
However, considering the poor applicability of the spin-wave theory
to one-dimensional antiferromagnets, the obtained description is
quite successful.
What is the difference between ferrimagnets and antiferromagnets, in
spite of the antiferromagnetic coupling between nearest neighbors in
common?
The surprising efficiency of the present spin-wave treatment can be
attributed to the ordered ground state \cite{Tian53} of ferrimagnets
and thus to the nondiverging sublattice magnetization.
The key constant ${\mit\Gamma}_1$ is nothing but the quantum spin
reduction
\begin{equation}
   \frac{1}{N}\sum_j\langle a_j^\dagger a_j\rangle_{\rm g}
  =\frac{1}{N}\sum_j\langle b_j^\dagger b_j\rangle_{\rm g}\,,
\end{equation}
where $\langle A\rangle_{\rm g}$ denotes the ground-state average of
$A$.
${\mit\Gamma}_1$ monotonically decreases as $S/s$ increases,
diverging at $S/s=1$ and vanishing for $S/s\rightarrow\infty$.
Since the spin reduction ${\mit\Gamma}_1$ can be a measure for the
validity of the spin-wave description, the spin-wave theory in
general works better as $S/s$, as well as $Ss$, increases.

   In this context, it is interesting to observe the ground-state
spin correlations,
\begin{eqnarray}
   &&
   f_S(r)=
   \left\{
   \begin{array}{ll}
      S_j^zS_{j+l}^z & \quad\mbox{for\ \ }r=2la    \,,\\
      S_j^zs_{j+l}^z & \quad\mbox{for\ \ }r=(2l+1)a\,,
   \end{array}
   \right.
   \label{E:fS}\\
   &&
   f_s(r)=
   \left\{
   \begin{array}{ll}
      s_j^zs_{j+l  }^z & \quad\mbox{for\ \ }r=2la    \,,\\
      s_j^zS_{j+l+1}^z & \quad\mbox{for\ \ }r=(2l+1)a\,.
   \end{array}
   \right.
   \label{E:fs}
\end{eqnarray}
\vskip 0mm
\begin{flushleft}
\mbox{\psfig{figure=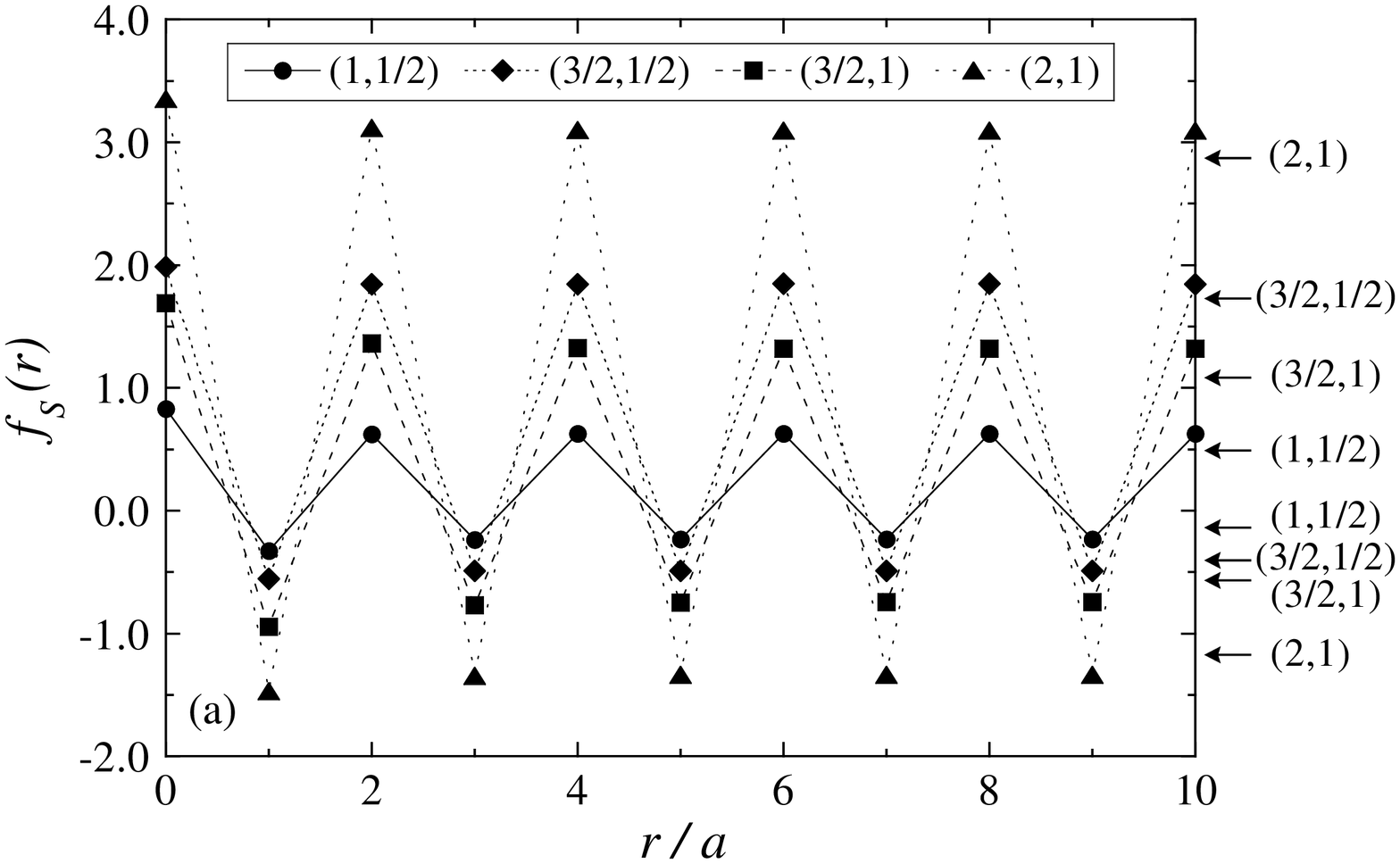,width=88mm,angle=0}}
\vskip 0mm
\mbox{\psfig{figure=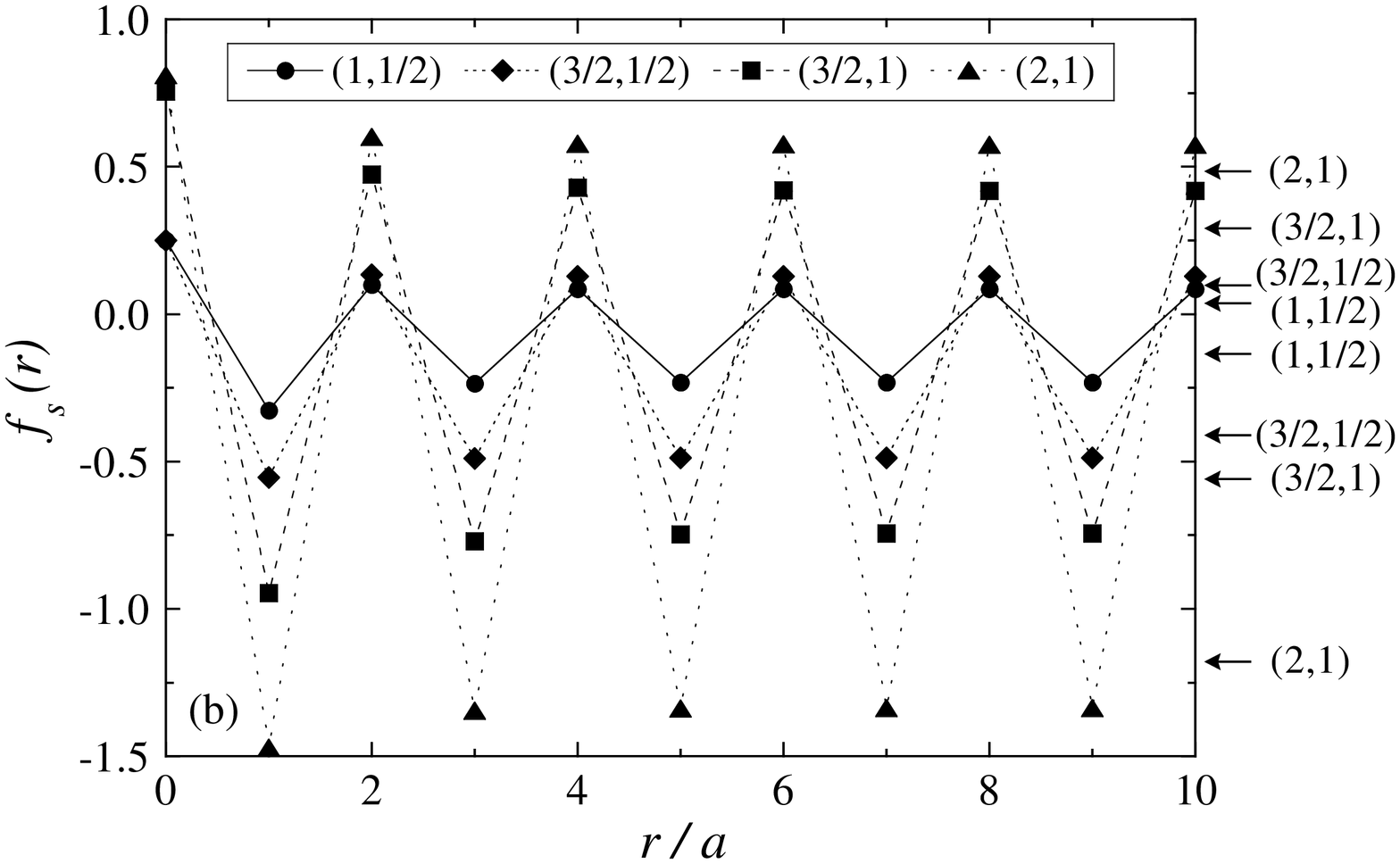,width=88mm,angle=0}}
\end{flushleft}
\begin{figure}
\vskip -2mm
\caption{Quantum Monte Carlo estimates of the longitudinal
         spin-correlation functions in the ground state with
         $M=(S-s)N$;
         (a) $f_S(r)$:
             correlations between a larger spin and any other spin;
         (b) $f_s(r)$:
             correlations between a smaller spin and any other spin.
         The arrows indicate the $r\rightarrow\infty$ asymptotic
         values obtained by the spin-wave theory.}
\label{F:Ss}
\end{figure}
Figure \ref{F:Ss} shows the small-$r$ initial behavior of $f_S(r)$
and $f_s(r)$, which demonstrates the considerably small correlation
length and the existence of the long-range order in the present
system.
For the asymptotic correlation between the two larger spins far
distant from each other, for example, the spin-wave estimate deviates
from the actual value by $23$\%, $7$\%, $18$\%, and $7$\% for
$(S,s)=(1,\frac{1}{2}),\,(\frac{3}{2},\frac{1}{2}),\,
       (\frac{3}{2},1)$, and $(2,1)$, respectively.
We find that the above-defined criterion in terms of $S/s$ and $Ss$
works fairly well.
The theory is not so bad even in the extreme quantum case.

   Quite recently, the nuclear spin relaxation time in a
one-dimensional ferrimagnetic Heisenberg model compound,
NiCu(C$_7$H$_6$N$_2$O$_6$)(H$_2$O)$_3$$\cdot$2H$_2$O,
has been measured \cite{Fuji} and its temperature and field
dependences have successfully been interpreted within the framework
of the spin-wave theory \cite{Yama}.
The theory must still be open to further applications to this
fascinating system.
We hope that the present study will motivate further explorations
into low-dimensional ferrimagnets in both theoretical and
experimental fields.
The ferromagnetic and antiferromagnetic mixed nature may further be
discussed from different points of view.
For instance, the topological excitations such as instantons in
ferromagnets and the topological terms in antiferromagnets might give
further support to the present understanding of ferrimagnets.

\acknowledgments

   This work was supported by the Japanese Ministry of Education,
Science, and Culture through Grant-in-Aid No. 11740206 and by the
Sanyo-Broadcasting Foundation for Science and Culture.
T. F. was supported by the JSPS Postdoctoral Fellowship for Research
Abroad.
The numerical computation was done in part using the facility of the
Supercomputer Center, Institute for Solid State Physics, University of
Tokyo.

\widetext
\begin{table}
\caption{Comparison between the linear-spin-wave (LSW),
         interacting-spin-wave (ISW), and numerical estimates of the
         ground-state energy $E_{\rm g}$, the antiferromagnetic
         excitation gap ${\mit\Delta}$, and the curvature of the
         ferromagnetic dispersion, $v$.
         The most accurate numerical estimates of $E_{\rm g}$ and
         ${\mit\Delta}$ are obtained by the exact diagonalization
         (Exact), whereas those of $v$ by the quantum Monte Carlo
         (QMC).}
\begin{tabular}{cccccccccc}
&
\multicolumn{3}{c}{$E_{\rm g}/NJ$} &
\multicolumn{3}{c}{${\mit\Delta}/J$} &
\multicolumn{3}{c}{$v$} \\
\cline{2-4}
\cline{5-7}
\cline{8-10}
{\raisebox{1.5ex}[0pt]{$(S,s)$}} &
LSW & ISW & Exact & LSW & ISW & Exact & LSW & ISW & QMC \\
\hline
$(1,\frac{1}{2})$ &
$-1.4365$ & $-1.4608$ & $-1.4541(1)$ &
$1$ & $1.6756$ & $1.759(1)$ &
$\frac{1}{2}$ & $0.3804$ & $0.37(1)$ \\
$(\frac{3}{2},\frac{1}{2}$) &
$-1.9580$ & $-1.9698$ & $-1.9672(1)$ &
$2$ & $2.8025$ & $2.842(1)$ &
$\frac{3}{8}$ & $0.3390$ & $0.31(1)$ \\
$(\frac{3}{2},1)$ &
$-3.8281$ & $-3.8676$ & $-3.861(1)$ &
$1$ & $1.5214$ & $1.615(5)$ &
$\frac{3}{2}$ & $1.1319$ & $0.90(3)$ \\
$(2,1)$ &
$-4.8729$ & $-4.8973$ & $-4.893(1)$ &
$2$ & $2.6756$ & $2.730(5)$ &
$1$ & $0.8804$ & $0.76(2)$ \\
\end{tabular}
\label{T:comp}
\end{table}

\end{document}